\documentclass[%
 reprint,
showpacs,
 amsmath,amssymb,
prb,
]{revtex4-1}
\usepackage{graphicx}% Include figure files
\usepackage{dcolumn}% Align table columns on decimal point
\usepackage{bm}% bold math
\usepackage{braket}
\usepackage{color}
\begin{document}
%\preprint{APS/123-QED}
%%%%%%%%%%%%%%%
\title{Stability of flat zero-energy states at the dirty surface of a 
nodal superconductor}
\author{Satoshi Ikegaya$^{1}$}
\email{satoshi-ikegaya@eng.hokudai.ac.jp}
\author{Yasuhiro Asano$^{1,2,3}$}%
\affiliation{$^{1}$Department of Applied Physics,
Hokkaido University, Sapporo 060-8628, Japan\\
$^{2}$Center of Topological Science and Technology,
Hokkaido University, Sapporo 060-8628, Japan\\
$^{3}$Moscow Institute of Physics and Technology, 141700 Dolgoprudny, Russia}
%%%%%%%%%%%%%%%
\date{\today}
%%%%%%%%%%%%%%%
\begin{abstract}
We discuss the stability of highly degenerate zero-energy states that
appear at the surface of a nodal superconductor preserving time-reversal symmetry.
The existence of such surface states is a direct consequence of 
the nontrivial topological numbers defined in the restricted Brillouin zones in the clean limit.
 In experiments, however, potential disorder is inevitable near the surface of a real superconductor,
which may lift the high degeneracy at zero energy. 
We show that an index defined in terms of the chiral eigenvalues of the zero-energy states
can be used to measure the degree of degeneracy at zero energy in the presence of potential disorder.
We also discuss the relationship between the index and the topological numbers.
\end{abstract}
%%%%%%%%%%%%%%%
\maketitle
%%%%%%%%%%%%%%%

\section{Introduction}

%-----(general interest)-----
The discovery of a topological insulator~\cite{kane_1,kane_2} had an impact on 
researchers studying the physics of superconductivity. 
The gapped band structures in a superconductor can also be topologically nontrivial.
The bulk-boundary correspondence,
which is the interrelationship between the nontrivial topological invariant in superconducting states in the bulk and
the number of gapless states at its surface,
immediately ensures the existence of surface bound states.
Recently such topologically nontrivial superconductors have attracted enormous attention
due to the existence of exotic surface bound states, some of which are composed of Majorana particles~\cite{Wiczek}.
Using the ten-fold classification as a basis~\cite{schnyder_0},
the early studies in this context were devoted to \textsl{fully gapped} topological superconductors.

%-----( nodal superconductors )-----
The ten-fold topological classification, however, covers only some real superconductors.
A number of real unconventional superconductors display nodes in the superconducting gap.
Nevertheless, such a \textsl{nodal} superconductor can often host highly degenerate surface bound states at 
zero energy (Fermi level). 
With of a spin-triplet superconductor, the 
degenerate surface zero-energy states (ZESs) are referred to as a Majorana flat band. 
The sign change in the pair potential on the Fermi surface, which is possible only in the presence 
of nodes in the gap functions, is the source of a topologically nontrivial superconducting phase 
preserving time-reversal symmetry~\cite{yt95,ya04}.
A prescription called dimensional reduction enables us to topologically characterize such 
nodal superconductors~\cite{sato_2,schnyder_5}.
In a $d$-dimensional superconductor, it is possible to choose a one-dimensional 
Brillouin zone by fixing a ($d$-1)-dimensional wave number at a certain point (say $k$).
When the energy spectra in the one-dimensional Brillouin zone at $k$ have gaps,
we can define the winding number $W(k)$~\cite{sato_2,schnyder_5}. 
According to the bulk-boundary correspondence for each Brillouin zone,
 a nodal superconductor often hosts degenerate ZESs at its clean surface. 
Namely, $\sum_{k}|W(k)|$-fold degenerate ZESs are expected 
at a surface parallel to $k$.
Actually the existence of degenerate surface ZESs has been suggested in time-reversal 
unconventional superconductors~\cite{buchholts,hu,yt95,ya04},
noncentrosymmetric superconductors~\cite{yt10,yada,schnyder_1,schnyder_2,schnyder_3},
semiconductor/superconductor heterostructures~\cite{alicea,you,law,si15},
superconductor/topological insulator heterostructures~\cite{deng}, and
superconducting Weyl semimetals~\cite{franz}.
It is widely accepted that 
flat ZESs cause various anomalies in low-energy transport such as the zero-bias 
anomaly in the conductance
of a normal-metal/superconductor junction~\cite{yt95,ya04,yt04,ya07,si15} 
and the fractional Josephson effect in a superconductor/insulator/superconductor junction
~\cite{yt96-1, barash, kwon, ya06,si16_2}.
These phenomena are unique to topologically nontrivial superconductors.

In experiments, however,
potential disorder is inevitable in the vicinity of the surface or junction interface 
of a superconductor. 
The one-dimensional Brillouin zone is ill-defined with the disordered potential
breaking the translational symmetry.
Therefore,
the winding number $W(k)$ can no longer use to  predict the number of 
ZESs at a dirty surface~\cite{schnyder_6}.
In other words, the potential disorder may lift the high degeneracy in the surface ZESs 
and may wash out the characteristic transport properties.
Such a situation requires a theoretical tool that mesures the stability of 
degenerate ZESs in the presence of potential disorder.
This paper addresses this issue and will provide experimentalists with helpful information.

By paying attention to the chiral symmetry of a Bogoliubov-de Gennes (BdG)
Hamiltonian~\cite{si15,tewari,niu,diez,si16},
we show that a mathematical index, $N_{\rm ZES}$, well characterizes
the number of ZESs at a dirty surface. 
The index $N_{\rm ZES}$ is an invariant defined in terms of
the chirality of the surface ZESs and is closely related to the one-dimensional 
winding number $W(k)$ ~\cite{sato_2}.
We conclude that the index $N_{\rm ZES}$ calculated in a clean superconductor exactly 
predicts the degree of degeneracy in ZESs at the dirty surface of a nodal superconductor.
Numerical simulations for several nodal superconductors ensure
the validity of the conclusion.

The organization of this paper is as follows.
In Sec.~\ref{sec:1},
we discuss the one-dimensional winding number
for a nodal superconductor preserving chiral symmetry.
The index $N_{\rm ZES}$ is defined in terms of the chiral eigenvalues of ZESs 
and is connected to the one-dimensional winding number through
the index theorem in mathematics. 
In Sec.~\ref{sec:2},
we confirm the validity of our conclusion for several superconductors 
such as $p$-, $d$- and $f$-wave unconventional superconductors, and
two noncentrosymmetric superconductors.
Section~\ref{sec:4} provides our conclusion.

%%%%%%%%%%
\section{Chiral symmetry and Index theorem}
\label{sec:1}
%%%%%%%%%%

\subsection{Winding number in a clean superconductor}
%-----( BdG Hamiltonian )-----
We begin our discussion with a brief summary of the topological property of a nodal superconductor in the clean limit.
The BdG Hamiltonian in momentum space is represented by,
\begin{gather}
H_0 (\boldsymbol{k}) = \left[
\begin{array}{cc}
\xi_0 (\boldsymbol{k}) & \Delta (\boldsymbol{k}) \\
- \Delta^{\ast} (-\boldsymbol{k}) & -\xi_0^{\ast} (-\boldsymbol{k}) \\
\end{array}\right], \label{eq:bdg_gene}
\end{gather}
where $\xi_0 (\boldsymbol{k})$ denotes the $M\times M$ Hamiltonian for an electron,
$\Delta (\boldsymbol{k})$ is the $M\times M$ pair potential and
where $M$ represents the number of degrees of freedom for an electron such as spin and 
band.
%-----( Symmetry )-----
The BdG Hamiltonian intrinsically preserves particle-hole symmetry
\begin{align}
&\Xi \; H_0 (\boldsymbol{k}) \; \Xi^{-1} =- H_0 (-\boldsymbol{k}),\\
&\Xi=  {\cal C} {\cal K}, \quad
{\cal C} = \left[
\begin{array}{cc}
0 & I \\
I & 0 \\
\end{array}\right] ,
\end{align}
where $I$ is the $M \times M$ unit matrix and
${\cal K}$ denotes the complex-conjugation operator.
We assume that the BdG Hamiltonian preserves 
time-reversal or time-reversal-like symmetry as
\begin{align}
&{\cal T}_{\pm} \; H_0 (\boldsymbol{k}) \; {\cal T}_{\pm}^{-1} =H_0 (-\boldsymbol{k}),\\
&{\cal T}_{\pm}=  {\cal U}_{\pm} {\cal K}, \quad
 {\cal U_{\pm}} = \left[
\begin{array}{cc}
u_{\pm} & 0 \\
0 & u_{\pm}^{\ast} \\
\end{array}\right] {\cal K},
\end{align}
%where $U$ is a unitary matrix and satisfies $UU^{\ast}=- I$.
where $u_{\pm}$ is an $M \times M$ unitary matrix satisfying  $u_{\pm}u_{\pm}^{\ast}=\pm I$.
Time-reversal symmetry is denoted with  $u_{-}u_{-}^{\ast}= -I$,
while time-reversal-like symmetry is denoted with $u_{+}u_{+}^{\ast}= I$.
By combining particle-hole and time-reversal symmetry,
it is possible to show the chiral symmetry of the BdG Hamiltonian,
\begin{align}
&\Gamma \; H_0 (\boldsymbol{k}) \; \Gamma^{-1} = - H_0 (\boldsymbol{k}), \label{chiral-}\\
&\Gamma= -i {\cal C} {\cal U}_- = \left[
\begin{array}{cc}
0 & -i u_-^{\ast} \\
-i u_- &  0 \\
\end{array}\right].
\end{align}
The chiral symmetry for a case of time-reversal-like symmetry is 
also defined in a similar way
\begin{align}
&\Gamma \; H_0 (\boldsymbol{k}) \; \Gamma^{-1} = - H_0 (\boldsymbol{k}), \label{chiral+} \\
&\Gamma= - {\cal C} {\cal U}_+ = \left[
\begin{array}{cc}
0 & -u_+^{\ast} \\
-u_+ &  0 \\
\end{array}\right].
\end{align}

%-----( One-dimensional winding number and the ZESs on the clean surface)-----
The pair potential under consideration has nodes. Namely 
$\Delta (\boldsymbol{k}_{\rm node})=0$ is satisfied 
at nodal points $\boldsymbol{k}_{\rm node}$ on the Fermi surface.
Therefore, it is impossible to characterize such superconducting states
topologically in terms of the wave function of the whole Brillouin zone.
Alternatively, we define a winding number in a one-dimensional 
Brillouin zone by fixing $\boldsymbol{k}_\parallel$ at a certain point. 
The momentum $k_\perp$ indicates a superconducting state 
in a one-dimensional Brillouin zone. 
The winding number is defined by~\cite{sato_2}
\begin{align}
W(\boldsymbol{k}_{\parallel})
= \frac{i}{4 \pi}
\int dk_{\perp} {\rm Tr}[
 \Gamma 
H_0^{-1} (\boldsymbol{k})
\partial_{k_{\perp}} H_0 (\boldsymbol{k}) ].
\label{eq:wind}
\end{align}
Since $\boldsymbol{k}_{\rm node}$ represents nodal points on the Fermi surface,
the relation $\xi (\boldsymbol{k}_{\rm node})=0$ holds simultaneously.
The winding number $W(\boldsymbol{k}_{\parallel})$
is ill-defined when the integration path along $k_\perp$ in Eq.~(\ref{eq:wind}) intersects
$\boldsymbol{k}_{\rm node}$. 
Therefore, we have to choose $\boldsymbol{k}_\parallel$ so that $k_\perp$ can be kept away from the nodal points. 
When $W(\boldsymbol{k}_{\parallel})$ is nonzero in a finite region of $\boldsymbol{k}_\parallel$, 
dispersionless ZESs with respect to $\boldsymbol{k}_{\parallel}$
appear at a clean surface parallel to $\boldsymbol{k}_{\parallel}$
~\cite{buchholts,hu,yt95,ya04, yt10,yada,schnyder_1,schnyder_2,schnyder_3,alicea,you,law,si15,deng,franz}. 
The number of ZESs at a \textsl{clean} surface is represented by
\begin{align}
N_{\rm clean} 
= {\sum_{\boldsymbol{k}_{\parallel}}}^{\prime} |W(\boldsymbol{k}_{\parallel})|,\label{n_clean}
\end{align}
where ${\sum_{\boldsymbol{k}_{\parallel}}}^{\prime}$ denotes
a summation over $\boldsymbol{k}_{\parallel}$ excluding the nodal points.
In what follows, we describe the degree of degeneracy in the ZESs
at a \textsl{dirty} surface in the presence of potential disorder.
The random impurity potential in the bulk region strongly suppresses the 
unconventional superconducting pair potential. 
Thus we consider the effects of 
the potential disorder only near a surface.

\subsection{Zero-energy states at a dirty surface}

%
%---------------------------------------------------------------------------
\begin{figure}[hhhh]
\begin{center}
\includegraphics[width=0.4\textwidth]{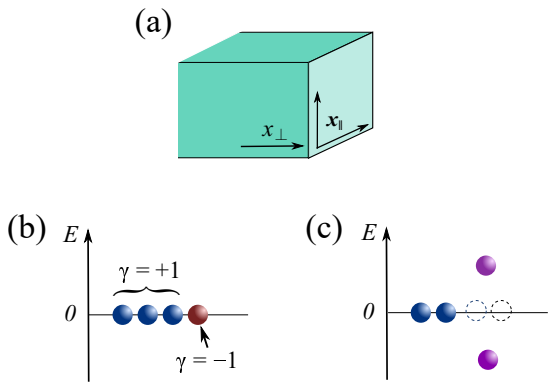}
\caption{(Color online) (a) Schematic image of
a semi-infinite nodal superconductor.
(b) Four-fold degenerate zero-energy states (ZESs) in the absence of random potential.
Three of them belong to the positive chiral eigenvalue, (i.e., $N_+=3$).
One remaining ZES belongs to the negative chiral eigenvalue (i.e., $N_-=1$).
The four-fold degeneracy is protected by translational symmetry. 
(c) In the presence of random potential, 
a positive and a negative chiral ZES form a pair and departs from zero energy.
However two positive chiral ZESs remain at zero energy.
The index $N_{\rm ZES}= N_+ - N_- =2$ represents the number of 
ZESs remaining at zero energy in the presence of random potential.
}
\label{fig:chiral}
\end{center}
\end{figure}
%------------------------------------------------------------------------
%

%-----( Chiral property )-----

We consider a semi-infinite superconductor that occupies 
$x_{\perp} \leq 0$ as shown in Fig.~\ref{fig:chiral} (a).
We apply the periodic boundary condition in a direction parallel to the surface $\boldsymbol{x}_\parallel$.
The BdG Hamiltonian in real space $H_0 (\boldsymbol{r})$ is obtained 
by replacing the momentum $\boldsymbol{k}$ by $-i \nabla_{\boldsymbol{r}}$.
The random impurity potential in the vicinity of the surface is represented by
\begin{align}
{V}_{\mathrm{imp}}(\boldsymbol{r})=\left[
\begin{array}{cc}
v (\boldsymbol{r}) I & 0 \\
0 & -v (\boldsymbol{r}) I \\
\end{array}\right],
\end{align}
where the random potential $v (\boldsymbol{r})$ disappears rapidly with increases in $x_{\perp}$ 
from the surface.
The total Hamiltonian is given by
\begin{align}
H(\boldsymbol{r}) = H_0(\boldsymbol{r}) + V_{\mathrm{imp}} (\boldsymbol{r}).
\end{align}
The momentum $\boldsymbol{k}_{\parallel}$ is no longer a good quantum number 
because the impurity potential breaks the translational symmetry.
As a result, it is impossible to define the one-dimensional winding 
number $W(\boldsymbol{k}_{\parallel})$ in the presence of the potential disorder.
%Moreover the random potential may hybridize the highly degenerate ZESs which 
%lifts the degeneracy at the zero-energy.
%Thus the degeneracy of the ZESs is considered to be fragile under the potential disorder.
However, the Hamiltonian 
$H(\boldsymbol{r})$ preserves the chiral symmetry of the Hamiltonian in Eqs.~(\ref{chiral-}) or (\ref{chiral+}), 
which is the most important factor in the argument below.

The central ingredient of our theory consists of
the following two important properties of the eigenstates in the presence of
chiral symmetry~\cite{sato_2}(i.e., $\left\{ H, \Gamma\right\}=0$).
\begin{itemize}
\item[(i)]
First, the zero-energy states of $H$ are
simultaneously the eigenstates of the chiral operator $\Gamma$.
Since $\Gamma^2 = 1$, the eigenvalue of $\Gamma$
 is either $\gamma=+1$ or $\gamma=-1$. 
Namely, a ZES satisfying
$
H_0 (\boldsymbol{r}) 
\varphi_{\gamma} (\boldsymbol{r}) = 0
$
also satisfies
$
\Gamma \varphi_{\gamma} (\boldsymbol{r}) 
= \gamma \varphi_{\gamma} (\boldsymbol{r}). 
$
We refer to $\gamma$ as the chirality in the following. 
\item[(ii)]
Second, the nonzero energy states of ${ H}$ are
described by the linear combination of the two states:
one has $\gamma= + 1$ and
the other has $\gamma= - 1$.
Namely, a nonzero energy state is described as
$
\varphi_{E \neq 0} (\boldsymbol{r})
= c_+ \chi_+(\boldsymbol{r}) + c_- \chi_-(\boldsymbol{r})
$, where 
$
\Gamma \chi_{\pm}(\boldsymbol{r})
= \pm \chi_{\pm}(\boldsymbol{r})
$.
Moreover, the relation $|c_+| = |c_-|$ always holds~\cite{si15}.
\end{itemize}

Here we define two integer numbers, $N_+$ and $N_-$, in the clean limit.
According to property (i),
we can immediately conclude that each ZES at a clean surface belongs to 
either the positive or the negative chiral state.
The integer $N_+$ ($N_-$) is the number of ZESs that have the positive (negative) chiral eigenvalue. 
[See also Fig.~\ref{fig:chiral} (b)].
The total number of ZESs at a clean surface is represented by $N_+ + N_-$, which must be identical to 
$N_{\rm clean}$ in Eq.~(\ref{n_clean}).

%-----( Chirality analysis and the index $N_{\rm ZES}$ )-----
The stability of the flat ZESs in the presence of impurities can be discussed by using 
property (ii). 
A ZES departs from zero energy only when it can form a pair with its chiral partner.
When $N_+ > N_-$, for example, $N_-$ negative chiral ZESs can couple to 
$N_-$ positive chiral ZESs under potential disorder. 
As a result, they form nonzero energy states whose number is $2 N_-$. 
However, $N_+ - N_-$ positive chiral states remain at zero 
energy even in the presence of impurities
because their chiral partner is absent.
The integer number defined by 
\begin{align}
N_{\mathrm{ZES}}=N_+ - N_-,
\end{align}
represents the number of ZESs that remain at a dirty surface.
When $N_+ < N_-$,
the number of ZESs at a dirty surface is given by $ N_- - N_+$.
Therefore, in general,
$|N_{\rm ZES}|$ is the degree of degeneracy at zero energy in the presence of potential disorder. 
The essence of this argument is illustrated in Figs.~\ref{fig:chiral} (b) and (c) with $N_+=3$ and $N_-=1$.
In Fig.~\ref{fig:chiral} (b), we consider four ZESs at a clean surface. 
In Fig.~\ref{fig:chiral} (c), we introduce the impurity potential at the surface.
Although the index $N_{\rm ZES}$ is defined in the presence of translational symmetry, 
it represents the degree of the degeneracy at zero energy in the absence of translational symmetry. 
This is the main conclusion of our paper.

%-----( Index theorem )-----
\subsection{Relation with topological number} 
At the end of this section, we discuss the topological aspect of $N_{\rm ZES}$.
As examined in Ref.~\onlinecite{sato_2},
the index theorem relates to the winding number $W(\boldsymbol{k}_{\parallel})$
and the number of ZESs on a clean surface as follows
\begin{align}
W(\boldsymbol{k}_{\parallel}) =
\pm \left[
n_+(\boldsymbol{k}_{\parallel}) - n_-(\boldsymbol{k}_{\parallel}) \right],
\label{eq:wind_chi_1}
\end{align}
where $n_+(\boldsymbol{k}_{\parallel})$ ($n_-(\boldsymbol{k}_{\parallel})$)
denotes the number of positive (negative) chiral zero-energy states at $\boldsymbol{k}_{\parallel}$.
There are two possible choice for the sign on the right-hand side of Eq.~(\ref{eq:wind_chi_1}).
When we consider the surface of a semi-infinite superconductor occupying $x_{\perp} \leq 0$
as shown in Fig.~\ref{fig:chiral}(a), we should choose the positive sign.
We should choose the positive sign at the
surface of a semi-infinite superconductor occupying $x_{\perp} \geq 0$
~\cite{sato_2}.
%When we choose the positive sign at a surface of a semi-infinite superconductor occupying $x_{\perp} \geq 0$
%as shown in Fig.~\ref{fig:chiral}(a), 
%we should choose the negative sign at a surface of a semi-infinite superconductor occupying $x_{\perp} \leq 0$
However, this sign has no physical meaning because
the number of ZESs is always given by $|n_+(\boldsymbol{k}_{\parallel}) - n_-(\boldsymbol{k}_{\parallel})|$.
As discussed in the previous subsection,
the index $N_{\rm ZES}$ is represented by the difference between the total numbers of
positive and negative chiral ZESs.
Therefore, by taking Eq.~(\ref{eq:wind_chi_1})  into account,
we find an important relation
\begin{align}
N_{\rm ZES} 
= {\sum_{\boldsymbol{k}_{\parallel}}}^{\prime} W(\boldsymbol{k}_{\parallel})
= \pm ( N_+ - N_-),
\label{eq:nzes}
\end{align}
as a result of the index theorem. 
More specifically,
the index $N_{\rm ZES}$ is a topological invariant defined in terms of the wave 
function in the superconducting states.
Simultaneously it is an invariant defined in terms of the zero-energy solutions in a differential equation. 
%Strictly speaking, there might be a possibility that
%more than $|N_{\rm ZES}|$ states remain at zero energy owing to the protection from any other symmetries.
The index theorem mathematically bridges the two different invariants.
In physics, the index $N_{\rm ZES}$ is an invariant that measures the degree of 
degeneracy of the ZESs staying at the dirty surface of a nodal superconductor. 
In the next section, we check the validity of our conclusion by performing numerical 
simulations on tight-binding model.

%%%%%%%%%%%%%%%
\section{Numerical results}
%%%%%%%%%%%%%%%
\label{sec:2}
%
%*****( Sentences to explain the objective of this subsection )*****
%
\subsection{unconventional superconductors}
We apply the general argument in Sec.~\ref{sec:1} to the 
several time-reversal superconductors in two-dimension.
The first example is the three types of unconventional 
superconductors characterized by $p_x$-, $d_{xy}$-, and $f$-wave pairing symmetry.
%
%*****( Sentences to explain the model )*****
%
We describe the present superconductors by
the $2 \times 2$ BdG Hamiltonian,
\begin{align}
&\hat{H}_{0} (\boldsymbol{k}) = \left[
\begin{array}{cc}
\xi(\boldsymbol{k}) & \Delta_{\mu} (\boldsymbol{k}) \\
\Delta_{\mu} (\boldsymbol{k}) & -\xi(\boldsymbol{k}) \\
\end{array}\right],\\
&\xi(\boldsymbol{k}) = \frac{\hbar^2 \boldsymbol{k}^2} {2m} - \mu_{\rm F},\\
&\Delta_{p_x} (\boldsymbol{k}) = \frac{\Delta_0} { k_{\rm F}} k_x,\\
&\Delta_{d_{xy}} (\boldsymbol{k}) =  \frac{\Delta_0} { k_{\rm F}^2} k_x k_y,\\
&\Delta_{f} (\boldsymbol{k}) =  \frac{\Delta_0} { k_{\rm F}^3} k_x ( k_{\rm F}^2 - 2 k_y^2 ),
\end{align}
where the subscript $\mu=p_x$, $d_{xy}$, $f$ labels the pairing symmetry,
$m$ denotes the effective mass of an electron,
$\mu_{\rm F}$ is the chemical potential,
$\Delta_0$ is the amplitude of the pair potential at zero temperature,
and $k_{\rm F}=\sqrt{2m \mu_{\rm F}}/\hbar$
represents the Fermi wave number.
The Hamiltonian satisfies
\begin{align}
\hat{\Gamma} \; \hat{H}_{0} (\boldsymbol{k})  \; \hat{\Gamma}^{-1}
= - \hat{H}_{0} (\boldsymbol{k}), 
\quad
\hat{\Gamma} = \left[
\begin{array}{cc}
0 & - i \\
 i & 0 \\
\end{array}\right],
\end{align}
which represents chiral symmetry of the Hamiltonian.

%
%*****( Sentences to obtain the $N_{\rm ZES}$ )*****
%
%-----( winding number )-----
%
%A superconductor spatially occupies $x\geq 0$.
%We consider bound states at the zero-energy appearing at a surface located at $x=0$.
The one-dimensional winding number in Eq.~(\ref{eq:wind})
can be further simplified to~\cite{sato_2, schnyder_5}
\begin{align}
W_{\mu} (k_y) = \frac{1}{2} \sum_{\xi(\boldsymbol{k}) =0}
\mathrm{sgn} [ \partial_{k_x} \xi(\boldsymbol{k}) ]
\mathrm{sgn} [ \Delta_{\mu} (\boldsymbol{k}) ]
\label{eq:wind_uncov}
\end{align}
where the summation is carried out for $k_x$ satisfying
$\xi(\boldsymbol{k})=0$ with fixed $k_y$.
From Eq.~(\ref{eq:wind_uncov}),
the winding number for each pairing symmetry
is calculated as
\begin{align}
&W_{p_x} (k_y)=
\left\{ \begin{array}{cl} 
1 & \text{for}\quad |k_y|< k_{\rm F} \\
0 & \text{for}\quad |k_y|> k_{\rm F},
\end{array}\right.
\label{eq:wind_p}\\
&W_{d_{xy}} (k_y)=
\left\{ \begin{array}{cl} 
1 & \text{for}\quad 0 <k_y< k_{\rm F} \\
-1 & \text{for}\quad 0>k_y>- k_{\rm F}\\
0 & \text{for}\quad |k_y|> k_{\rm F},
\end{array}\right.\\
&W_{f} (k_y)=
\left\{ \begin{array}{cl} 
1 & \text{for}\quad |k_y|< k_f \\
-1 & \text{for}\quad k_f < |k_y| <k_{\rm F}\\
0 & \text{for}\quad |k_y|> k_{\rm F}, \\
\end{array}\right.
\label{eq:wind_f}
\end{align}
where $k_f = k_{\rm F}/\sqrt{2}$.
The total number of topologically protected ZESs at the clean surface is calculated as
\begin{align}
N_{\rm clean} = 
{\sum_{k_y}}^{\prime}  | W_{\mu} (k_y) |,
\end{align}
according to the bulk-boundary correspondence.
%
%-----( disordered edge )-----
%
The number of ZESs at the dirty surface, on the other hand, 
 is evaluated by the index $N_{\rm ZES}$.
By substituting Eq.~(\ref{eq:wind_p}) - (\ref{eq:wind_f}) into Eq.~(\ref{eq:nzes}),
we obtain the index $N_{\rm ZES}$ for each pairing symmetry as
\begin{align}
|N_{\rm ZES}| =
\left\{ \begin{array}{cl} 
\sum_{|k_y|<k_{\rm F}} = N_{\rm clean} & \text{for $p_x$-wave} \\
0 & \text{for $d_{xy}$-wave}\\
\sum_{|k_y|<k_f}- \sum_{k_f<|k_y|<k_{\rm F}} \neq 0 & \text{for $f$-wave}.\\
\end{array}\right. \label{nzes_uc}
\end{align}

%
%*****( Sentences to explain the numerical calculation )*****
%---------------------------------------------------------------------------
\begin{figure}[hhhh]
\begin{center}
\includegraphics[width=0.4\textwidth]{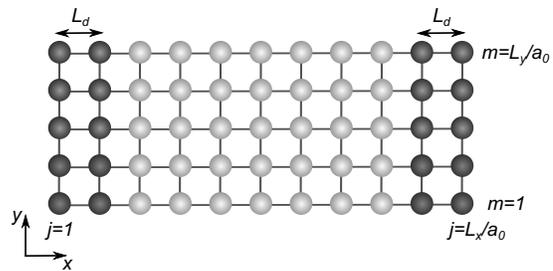}
\caption{(Color online)
Schematic picture of a superconductor on the tight-binding lattice. 
}
\label{fig:tb_model}
\end{center}
\end{figure}
%------------------------------------------------------------------------
%
%
%------------------------------------------------------------------------
\begin{figure*}
\includegraphics[width=0.95\textwidth]{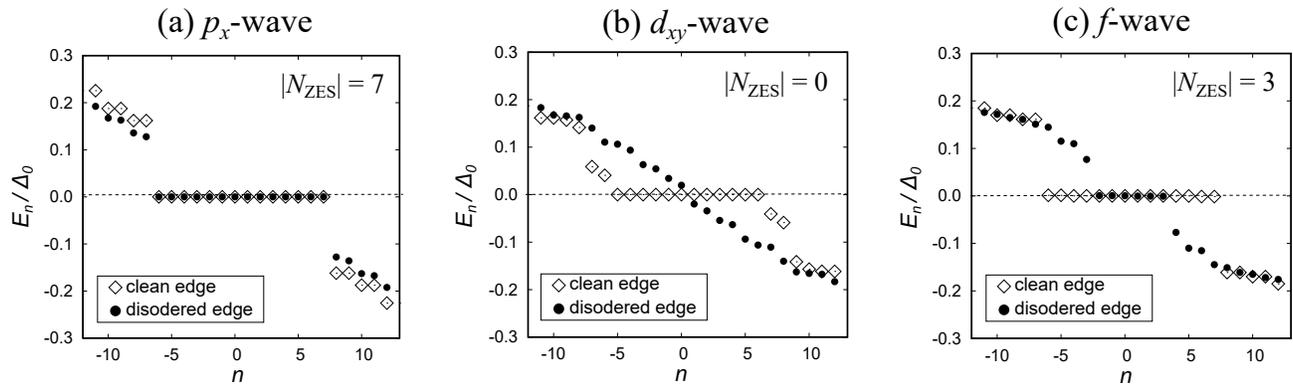}
\caption{Energy eigenvalues of (a) $p_x$- , (b)$d_{xy}$-, and (c) $f$-wave superconductor are plotted.
In numerical simulation, eigenvalues are calculated in decreasing order labeled by $n$ in the horizontal axis.
In the clean limit, as shown with the open symbols,
we find $2 \times N_{\rm clean} =14$ zero-energy states (ZESs) for $p_x$- and $f$-wave cases.
For a $d_{xy}$-symmetry,
the two states at a gap nodal point $k_y=0$ are leave from zero energy due to the finite size effect.
Thus the number of ZESs becomes $2 \times N_{\rm clean} =12$.
In the presence of potential disorder at two surfaces, as shown with the filled symbols,
the number of the ZESs is identical to $2 \times |N_{\rm ZES}|$ which is 14 in (a), 0 in (b), and 6 in (c).
}
\label{fig:eng_uncov}
\end{figure*}
%------------------------------------------------------------------------
To check the validity of Eq.~(\ref{nzes_uc}),
we numerically calculate the eigen energy of an 
isolating unconventional superconductor
on the two-dimensional tight-binding model as shown in Fig.~\ref{fig:tb_model}.
A lattice site is indicated by a vector
$\boldsymbol{r}= j \; a_0 \boldsymbol{x} + m\; a_0 \boldsymbol{y}$,
where $a_0$ denotes the lattice constant and
$\boldsymbol{x}$ ($\boldsymbol{y}$)
is a unit vector in the $x$ ($y$) direction.
The number of the lattice site in the $x$ and $y$
direction are denoted by $L_x/a_0$  and $L_y/a_0$, respectively.
In the $y$ direction,
the periodic boundary condition is applied.
In the $x$ direction,
we apply the hard-wall boundary condition.
When the sample length $L_x$ is much larger than the superconducting coherent 
length $\xi_0 = (\hbar^2 k_{\rm F}/m \Delta_0)$,
all the bound states in the vicinity of both $j=1$ and $j=L_x$ energetically localize at zero energy.
We introduce the impurity potential by adding the random on-site potentials
$v(\boldsymbol{r})$ in the outermost $L_d/a_0$ layers in the $x$-direction as shown in Fig.~\ref{fig:tb_model}.
The amplitude of $v(\boldsymbol{r})$ is given randomly 
in the range of $-V_{\rm I}/2 \leq v(\boldsymbol{r}) \leq V_{\rm I}/2$.
We numerically diagonalize 
the BdG Hamiltonian on the tight-binding model which is shown in Appendix~\ref{sec: tb_1}.
We fix several parameters as
$\mu_{\rm F}=1.5t$, $\Delta_0=1.0t$, $L_x=60a_0$ and $L_y=18a_0$
where $t$ denotes the nearest-neighbor hopping integral.
This parameter choice leads to $N_{\rm clean} = 7$ for a $p_x$- and a $f$-wave superconductor,
and $N_{\rm clean} = 6$ for a $d_{xy}$-wave superconductor.
The superconducting coherent length satisfies $\xi_0 \approx 2 a_0\ll L_x$.
The indexes $N_{\rm ZES}$ become $7$, $0$, and $3$ for
 $p_{x}$-, $d_{xy}$-, and $f$-wave pairing symmetry, respectively.
%
%*****( Sentences to confirm the validity of the N_{\rm ZES})*****
%-----( emphasize the result for the $f$-wave superconductor )----- 
%
In Figs.~\ref{fig:eng_uncov}(a) - \ref{fig:eng_uncov}(c),
we show the numerical results of energy eigenvalues,
where the eigen energy is labeled by an integer $n$.
The open symbols and the filled symbols respectively 
denote the energy eigenvalues in a superconductor with clean surfaces and those with dirty surfaces.
%The clean edge is described with $v(\boldsymbol{r})=0$.
We chose $V_{\rm I}=3.0t$ and $L_d=5a_0$ to realize the dirty surfaces.
As shown in Fig.~\ref{fig:eng_uncov}(c),
for instance, a $f$-wave superconductor with clean surfaces
has $2 \times N_{\rm clean}$ ($=14$) zero-energy states,
where the factor $2$ is derived from the contribution from two different surfaces in the $x$ direction.
In the presence of the impurity potentials, eight ZESs leave away from the zero-energy
as shown in the filled symbols, whereas six ZESs still keep staying at zero energy.
Since $2 \times |N_{\rm ZES}|=6$ under the present parameter choice, the argument  
in Sec.~II predicts the number of ZESs at a dirty surface exactly.  
Figures~\ref{fig:eng_uncov}(a) and \ref{fig:eng_uncov}(b) show the perfect agreement 
between our theory and numerical results.
 For a $p_x$-wave superconductor, $2 \times |N_{\rm ZES}|=14$ states remain at zero energy.
In a $d_{xy}$-wave case, ZESs are absent at the dirty surfaces.

%%%%%%%%%%%%%%%
\subsection{Noncentrosymmetric  superconductor I}
%\label{sec:3}
%%%%%%%%%%%%%%%
%
%*****( Sentences to explain the objective of this subsection )*****
%
Secondly we apply the argument in Sec.~\ref{sec:1} to 
the noncentrosymmetric superconductors (NCSs) in two-dimension.
%
%*****( Sentences to explain the model in general )*****
%
The BdG Hamiltonian for a NCS is given by
\begin{align}
&\check{H} (\boldsymbol{k}) = \left[
\begin{array}{cc}
\hat{h} (\boldsymbol{k}) & \hat{\Delta} (\boldsymbol{k}) \\
- \hat{\Delta} ^{\ast}(-\boldsymbol{k}) & -\hat{h}^{\ast} (-\boldsymbol{k}) \\
\end{array}\right],
\label{eq:bdgncs}\\
&\hat{h} (\boldsymbol{k})=\xi(\boldsymbol{k}) \hat{\sigma}_0
+ \boldsymbol{g}(\boldsymbol{k}) \cdot \hat{\boldsymbol{\sigma}},\\
&\hat{\Delta} (\boldsymbol{k})=i  [\psi(\boldsymbol{k})
+ \boldsymbol{d}(\boldsymbol{k}) \cdot \hat{\boldsymbol{\sigma}} ] \hat{\sigma}_y,
\end{align}
where $\hat{\boldsymbol{\sigma}}=(\hat{\sigma}_x,\hat{\sigma}_y,\hat{\sigma}_z)$ and $\hat{\sigma}_0$
denote Pauli matrices in spin space and the $2 \times 2$ unit matrix, respectively.
The absence of inversion symmetry leads the spin-orbit coupling (SOC) potential
denoted by $\boldsymbol{g}(\boldsymbol{k}) = - \boldsymbol{g}(- \boldsymbol{k})$.
Furthermore,
the pair potential becomes the admixture of
the even-parity spin-singlet component $\psi(\boldsymbol{k})=\psi(-\boldsymbol{k})$
and the odd-parity spin-triplet pair component $\boldsymbol{d}(\boldsymbol{k})=-\boldsymbol{d}(-\boldsymbol{k})$
because parity is no longer a good quantum index~\cite{sigrist,fujimoto}.
The spin-triplet pairing vector $\boldsymbol{d}(\boldsymbol{k})$ is set to be 
parallel to the polarization vector of the SOC~\cite{sigrist} (i.e., 
$\boldsymbol{d}(\boldsymbol{k}) \parallel \boldsymbol{g}(\boldsymbol{k})$). 
The BdG Hamiltonian satisfies
\begin{align}
\check{\Gamma} \; \check{H}_0 (\boldsymbol{k})  \; \check{\Gamma}^{-1} = - \check{H}_0 (\boldsymbol{k}),
\label{eq:chiral_ncs}
\quad
\check{\Gamma} = \left[
\begin{array}{cc}
0 & \hat{\sigma}_y \\
 \hat{\sigma}_y & 0 \\
\end{array}\right],
\end{align}
which represents the chiral symmetry of the Hamiltonian.

\begin{figure}[bbbb]
\begin{center}
\includegraphics[width=0.25\textwidth]{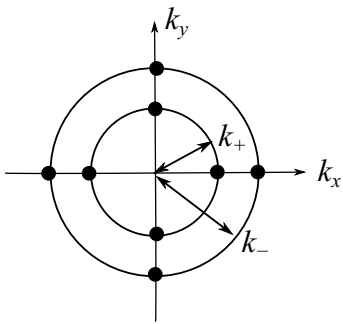}
\caption{Two Fermi surfaces under the Rashba SOC are illustrated.
The eight nodal points are indicated by the black dots.}
\label{fig:fermi_dxyp}
\end{center}
\end{figure}

%******************
%\subsection{Rashba spin-orbit coupling}
%\label{sec:3a}
%******************
%
%*****( Sentences to explain the model of the $(d_{xy} + p)$-wave superconductor )*****
%
A superconductor with ($d_{xy}+p$)-wave pairing symmetry is an example of NCS 
which host the flat ZES at its clean surface~\cite{yt10, yada,schnyder_6}.
Under the Rashba type SOC 
$\boldsymbol{g}_{\rm r}(\boldsymbol{k}) = 
\alpha( k_y \boldsymbol{x} - k_x \boldsymbol{y})$ with $\alpha$ being the coupling amplitude,
the normal state Fermi surface splits into the two
 circles as illustrated in Fig.~\ref{fig:fermi_dxyp}, where
the two wave numbers 
\begin{align}
&k_\pm = \mp \frac{m\alpha}{\hbar^2} + \sqrt{k_{\rm F}^2 +\left( \frac{m\alpha}{\hbar^2} \right)^2 },
%\\
%&k_- = + \frac{m\alpha}{\hbar^2} + \sqrt{k_{\rm F}^2 +\left( \frac{m\alpha}{\hbar^2} \right)^2 },
\end{align}
characterize the two Fermi surfaces.
The pair potential of the ($d_{xy}+p$)-wave superconductor is given as~\cite{yt10, yada,yada_2}
\begin{align}
\psi(\boldsymbol{k}) = \Delta_s f(\boldsymbol{k}),
\quad
\boldsymbol{d}(\boldsymbol{k})
= \Delta_t f(\boldsymbol{k}) \frac{\boldsymbol{g}_{\rm r}(\boldsymbol{k})}{\alpha k},
\end{align}
with $f(\boldsymbol{k}) = (k_x k_y/k^2)$ and $k = \sqrt{k_x^2 + k_y^2 }$.
In this pair potential, there are eight nodal points
which are located at $(\pm k_\pm, 0)$ and $(0, \pm k_\pm)$
 as illustrated in Fig.~\ref{fig:fermi_dxyp}.

%----------------------------------------------------------
%----------------------------------------------------------

%
%*****( Sentences to obtain the $N_{\rm ZES}$ )*****
%
%
%-----( winding number )-----
%
%Now we focus on the ZESs appearing at an edge in the $x$-direction.
By applying a unitary transformation shown in Appendix~\ref{sec:uni1},
it is possible to deform the BdG Hamiltonian of the  ($d_{xy}+p$)-wave superconductor as
\begin{align}
&\check{H}^{\prime}_0 (\boldsymbol{k})= \left[
\begin{array}{cc}
\hat{H}_+ (\boldsymbol{k}) & 0 \\
0 & \hat{H}_- (\boldsymbol{k}) \\
\end{array}\right],\\
&\hat{H}_{\pm}(\boldsymbol{k})= \left[
\begin{array}{cc}
\xi_{\pm} (\boldsymbol{k}) & -\Delta_{\pm}(\boldsymbol{k})\\
-\Delta_{\pm}(\boldsymbol{k}) &  - \xi_{\pm}(\boldsymbol{k}) \\
\end{array}\right],
\label{eq:ham_dp} \\
&\xi_{\pm}(\boldsymbol{k}) = \xi(\boldsymbol{k}) \pm |\boldsymbol{g}_{\rm r}(\boldsymbol{k})|,
\\
&\Delta_{\pm} (\boldsymbol{k}) =  f(\boldsymbol{k}) \left[\Delta_t \pm \Delta_s \right].
\end{align} 
The chiral symmetry in this new basis is represented as
\begin{align}
\hat{\Gamma}_{\pm} \; \hat{H}_{\pm} (\boldsymbol{k})  \; \hat{\Gamma}_{\pm}^{-1}
= - \hat{H}_{\pm} (\boldsymbol{k}), \quad
\hat{\Gamma}_{\pm} = \mp\hat{\sigma}_y.
\label{eq:chiral_dp}
\end{align}  
By using Eqs.~(\ref{eq:ham_dp}) and (\ref{eq:chiral_dp}),
the relevant winding number can be calculated as
~\cite{sato_2, schnyder_5}  
\begin{align}
&W (k_y) = W_{+} (k_y) -W_{-} (k_y),
\label{eq:wind_dxyp}\\
&W_{\pm} (k_y) = \frac{1}{2} \sum_{\xi_{\pm}(\boldsymbol{k}) =0}
\mathrm{sgn} [ \partial_{k_x} \xi_{\pm}(\boldsymbol{k}) ]
\mathrm{sgn} [ \Delta_{\pm} (\boldsymbol{k}) ],
\end{align}
where the summation is carried out for $k_x$ that satisfies
$\xi_{\pm}(\boldsymbol{k})=0$ with fixed $k_y$.  
From Eq.~(\ref{eq:wind_dxyp}), we obtain
\begin{align}
W(k_y)=
\left\{ \begin{array}{cl} 
2 \mathrm{sgn}(k_y) & \text{for}\quad |k_y|< k_+ \\
\mathrm{sgn} (k_y) & \text{for}\quad k_+ < |k_y| <k_-\\
0 & \text{for}\quad |k_y|> k_-, \\
\end{array}\right.
\end{align}
for $\Delta_s > \Delta_t$, and
\begin{align}
W(k_y)=
\left\{ \begin{array}{cl} 
0 & \text{for}\quad |k_y|< k_+ \\
- \mathrm{sgn} (k_y) & \text{for}\quad k_+ < |k_y| <k_-\\
0 & \text{for}\quad |k_y|> k_-, \\
\end{array}\right.
\end{align}
for $\Delta_t > \Delta_s$.
%
%-----( $N_{\rm ZES}$ )-----
%
The index $N_{\rm ZES}$ is calculated from Eq.~(\ref{eq:nzes}).
Since the winding number satisfies $W(k_y) = - W(-k_y)$,
we immediately find $N_{\rm ZES}=0$
for both $\Delta_s > \Delta_t$ and $\Delta_t > \Delta_s$.
%
%*****( Sentences to demonstrate the confirmation of $N_{\rm ZES}$ )*****
%

\begin{figure}[hhhh]
\begin{center}
\includegraphics[width=0.3\textwidth]{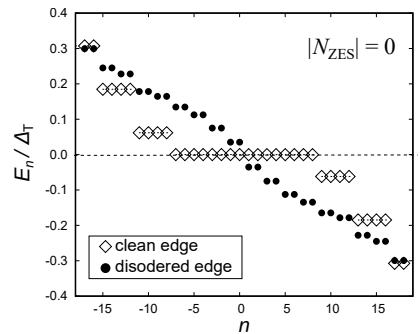}
\caption{Energy eigenvalues of a ($d_{xy}+p$)-wave superconductor
are plotted in the same manner as in Fig.~\ref{fig:eng_uncov}.
A energy is normalized by $\Delta_{\rm T} = \Delta_s + \Delta_t$.
In the clean limit, there are 16 zero-energy states as shown with the open symbols.
In the presence of random potential at the two surfaces, on the other hand,
the ZESs are absent as shown with the filled symbols in agreement 
with $N_{\rm ZES}=0$.
}
\label{fig:eng_dxyp}
\end{center}
\end{figure}

The figure~\ref{fig:eng_dxyp}
shows the eigenvalues of the BdG Hamiltonian for a ($d_{xy}+p$)-wave superconductor.
The expression of the Hamiltonian on the tight-binding model is given in Appendix ~\ref{sec: tb_2}.
We chose parameters as $\mu_{\rm F}=2.0t$, $\alpha=0.1t$,
$\Delta_s=0.8t$, $\Delta_t=0.2t$, $L_x=50a_0$, $L_y=10a_0$ and $L_d=5a_0$.
The open and the filled symbols denote the energy eigenvalues of a superconductor 
with the clean surface ($V_{\rm I}=0$) and the dirty surfaces ($V_{\rm I}=3.0$t), respectively.
We find the sixteen ZESs reflecting the nonzero winding numbers in the clean case.
The random potential at the surfaces completely lifts the degeneracy at zero energy as shown with 
the filled symbols. The numerical results agree with the argument in Sec.~II.
Since $N_{\rm ZES}=0$, the flat ZES in a ($d_{xy}+p$)-wave superconductor is fragile under the potential disorder.
At finte energies, the eigenvalues for dirty surfaces are always doubly degenerate, which corresponds 
to the Kramers doublets protected by time-reversal symmetry.

%---------------------------------------------------------------------------
%------------------------------------------------------------------------

\subsection{Noncentrosymmetric superconductor II}
\label{sec:3b}

\begin{figure}[tttt]
\begin{center}
\includegraphics[width=0.25\textwidth]{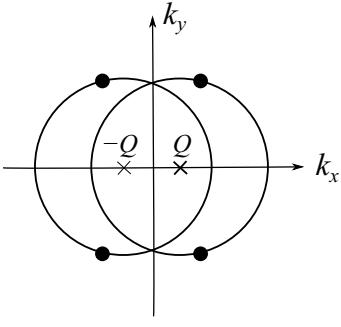}
\caption{Fermi surfaces under the Dresselhaus[110] SOC are illustrated.
The eight nodal points are indicated by the black dots.
Strictly speaking, the positions of the nodal
points depend on the parameters as shown in Eq.~(\ref{eq:node_sp}).
}
\label{fig:fermi_sp}
\end{center}
\end{figure}

In a zincblende semiconductor quantum well confined in the [110] crystal direction, 
the Dresselhaus [110] type spin-orbit coupling described by
$\boldsymbol{g}_{\rm d}(\boldsymbol{k}) = \beta k_x \boldsymbol{z}$
 becomes dominant.
The Hamiltonian for the normal states is given by
\begin{align}
\hat{h}_{\rm P} (\boldsymbol{k})
= \xi(\boldsymbol{k}) \hat{\sigma}_0 + \beta k_x \hat{\sigma}_z.
\label{eq:psh}
\end{align}
The electronic states described by Eq.~(\ref{eq:psh}) 
have been well studied in spintronics
because they show an unusual spin property called persistent spin helix~\cite{ bernevig, chang, kohda,salis}.
As shown in Appendix~\ref{sec:psh_rd},
the persistent spin-helix states can be also obtained in the thin film growing along the [001] crystal direction~\cite{ bernevig,chang}.
In what follows, we discuss the flat ZESs appearing at a surface of a proximity-induced superconducting
Dresselhaus[110] thin film described by
\begin{align}
&\check{H}_{0} (\boldsymbol{k}) = \left[
\begin{array}{cc}
\hat{h}_{\rm P} (\boldsymbol{k}) & \hat{\Delta}_{\rm P} (\boldsymbol{k}) \\
- \hat{\Delta}_{\rm P} ^{\ast}(-\boldsymbol{k}) & -\hat{h}_{\rm P}^{\ast} (-\boldsymbol{k}) \\
\end{array}\right],
\label{eq:bdgncs2}\\
&\hat{\Delta}_{\rm P}=i  \left[\Delta_s
+ \Delta_t\frac{k_x}{k_{\rm F}}  \hat{\sigma}_z\right] \hat{\sigma}_y,
\label{eq:s+p}
\end{align}
where we assume the $s$-wave pairing symmetry for the spin-singlet component.
The Dresselhaus[110] SOC potential shifts
the Fermi surfaces in the $k_x$ direction as illustrated in Fig.~\ref{fig:fermi_sp}.
The center of the Fermi surfaces are located at
$(\pm Q,0)$ with $Q = m \beta/\hbar^2$.
The superconducting gap has four nodes on the Fermi surface when the condition
\begin{align}
\frac{\beta k_{\rm F}}{\mu_{\rm F}} > \frac{\Delta_s^2-\Delta_t^2}{\Delta_s\Delta_t}
\label{eq:node}
\end{align}
is satisfied. 
The four nodal points are located at
\begin{align}
&(\pm r_{st}  k_{\rm F}, \pm k_{\rm Q}),
\label{eq:node_sp}\\
&k_{\rm Q} = k_{\rm F} \sqrt{1-  r_{st}^2 + r_{st}(\beta k_{\textrm{F}}/\mu_{\mathrm{F}})   }, 
\end{align}
as indicated by filled circle in Fig.~\ref{fig:fermi_sp},
where $r_{st} = (\Delta_s/\Delta_t)$.
%
%
%----------------------------------------------------------
%----------------------------------------------------------
%
%
The BdG Hamiltonian preserves both time-reversal and time-reversal-like symmetry as
\begin{align}
&\check{T}_{\pm} \; \check{H}_{0} (\boldsymbol{k}) \; \check{T}_{\pm}^{-1} = \check{H}_{0}(-\boldsymbol{k}),\\
&\check{T}_+ = \left[
\begin{array}{cc}
i \hat{\sigma}_x  & 0 \\
0 & -i \hat{\sigma}_x \\
\end{array}\right] {\cal K}, \quad
\check{T}_- = \left[
\begin{array}{cc}
i \hat{\sigma}_y  & 0 \\
0 & i \hat{\sigma}_y \\
\end{array}\right] {\cal K},
\end{align}
where $\check{T}_{\pm}^2=\pm 1$.
Therefore we obtain two different chiral symmetry operators as
\begin{align}
&\check{\Gamma}_{\pm} \; \check{H}_{0} (\boldsymbol{k}) \; \check{\Gamma}_{\pm}^{-1} = - \check{H}_{0}(\boldsymbol{k}),\\
&\check{\Gamma}_+ = \left[
\begin{array}{cc}
  0 & - i \hat{\sigma}_x \\
i \hat{\sigma}_x  & 0 \\
\end{array}\right] , \quad
\check{\Gamma}_- = \left[
\begin{array}{cc}
 0 & \hat{\sigma}_y \\
\hat{\sigma}_y & 0 \\
\end{array}\right].
\end{align}
By applying the unitary transformation as
\begin{align}
\check{H}^{\prime}(\boldsymbol{k}) = \check{U}_{0}^\dagger H_{0}(\boldsymbol{k}) \check{U}_{0}, \quad
\check{U}_{0}=\left[ \begin{array}{cccc}
1 & 0 & 0 & 0\\
0 & 0 & 1 & 0 \\
0 & 0 & 0 & 1 \\
0 & 1 & 0 & 0
\end{array}\right],
\end{align}
the BdG hamiltonian is block-diagonalized into the two $2 \times 2$ sectors as 
$\check{H}^\prime= \textrm{diag}(\hat{H}_1,\hat{H}_2)$,
\begin{align}
&\hat{H}_{j}(\boldsymbol{k})= \left[
\begin{array}{cc}
\xi_{j} (\boldsymbol{k}) & -\Delta_{j}(\boldsymbol{k})\\
-\Delta_{j}(\boldsymbol{k}) &  - \xi_{j}(\boldsymbol{k}) \\
\end{array}\right],\label{eq:ham_sp}
\end{align}
for $j=1-2$ with
\begin{align}
&\xi_{1(2)}(\boldsymbol{k}) = \xi(\boldsymbol{k}) + (-) \beta k_x,\\
&\Delta_{1(2)}(\boldsymbol{k}) = \left[\Delta_t\frac{k_x}{k_{\rm F}} + (-) \Delta_s \right].
\end{align}
The chiral symmetry of each block component is represented as
\begin{align}
&\hat{\Gamma}_{\pm, j} \; \hat{H}_{j} (\boldsymbol{k})  \; \hat{\Gamma}_{\pm, j}^{-1}
= - \hat{H}_{j} (\boldsymbol{k}),
\end{align}
for $j=1-2$, where
\begin{align}
&\hat{\Gamma}_{+, 1} = \hat{\sigma}_y,\quad \hat{\Gamma}_{+, 2}= \hat{\sigma}_y,
\label{eq:chiral_spp}
\end{align}
is originated from time-reversal like symmetry, and
\begin{align}
\hat{\Gamma}_{-, 1} = \hat{\sigma}_y, \quad \hat{\Gamma}_{-, 2} = -\hat{\sigma}_y.
\label{eq:chiral_spm}
\end{align}
is originated from time-reversal symmetry.
The definition of the winding number depends on the form of the chiral symmetry operator.
From the chiral symmetry operator originated from the time-reversal symmetry in Eq.~(\ref{eq:chiral_spm}),
the winding number is given by
\begin{align}
W(k_y) =& W_{1} (k_y) - W_{2} (k_y), \label{eq:wind4_m}
\end{align}
with
\begin{align}
W_j(k_y) =&\frac{1}{2} \sum_{\xi_{j}(\boldsymbol{k}) =0}
\mathrm{sgn} [ \partial_{k_x} \xi_{j}(\boldsymbol{k}) ]
\mathrm{sgn} [ \Delta_{j} (\boldsymbol{k}) ],
\end{align}
for $j=1-2$, where the summation is carried out for $k_x$ satisfying
$\xi_{j}(\boldsymbol{k})=0$ at a fixed $k_y$. 
The winding number in each sector is calculated to be
\begin{align}
W_1 (k_y)= W_2 (k_y)=
\left\{ \begin{array}{cl} 
1 & \text{for}\quad |k_y|< k_{\rm Q} \\
0 & \text{for}\quad |k_y|> k_{\rm Q}.
\end{array}\right.,
\label{eq:wind_sp}
\end{align}
Although the winding number in each sector $W_{1(2)} (k_y)$ is nontrivial,
the relation $W(k_y)=0$ always holds.
As a consequence, we find $N_{\mathrm{ZES}}=0$.
The results suggest the degeneracy at zero energy would be fragile under the potential disorder.
{However, the winding number originated from time-reversal-like symmetry can be nontrivial 
because the winding number defined with Eq.~(\ref{eq:chiral_spp}) is given as}
\begin{align}
W(k_y) = W_{1} (k_y) + W_{2} (k_y). \label{eq:wind4_p}
\end{align}
We find 
\begin{align}
W (k_y)=&
\left\{ \begin{array}{cl} 
2 & \text{for}\quad |k_y|< k_{\rm Q} \\
0 & \text{for}\quad |k_y|> k_{\rm Q}.
\end{array}\right.,
\label{eq:wind_sp}\\
N_{\rm clean} =& N_{\rm ZES}= 
2 \sum_{|k_y|< k_{\rm Q}}.
\end{align}

\begin{figure}[hhhh]
\begin{center}
\includegraphics[width=0.3\textwidth]{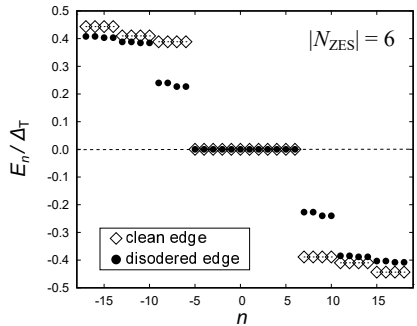}
\caption{Energy eigenvalues of a
NCS with Dresselhaus[110] SOC are plotted in the same manner as Fig.~\ref{fig:eng_uncov}.
The results are normalized by $\Delta_{\rm T} = \Delta_s + \Delta_t$.
The number of ZESs is 12 in the clean case as shown with the open symbols.
All of the ZESs keep staying at zero energy even in the presence of random potential 
at the two surfaces as predicted by the index $2 \times |N_{\rm ZES}|=12$.
}
\label{fig:eng_sp}
\end{center}
\end{figure}

In Fig.~\ref{fig:eng_sp}, 
we show the energy eigenvalues of the NCS with the Dresselhaus[110] SOC 
on the tight-binding model.
The BdG Hamiltonian used in the numerical simulation is show in Appendix~\ref{sec: tb_3}.
We chose parameters as $\mu_{\rm F} =1.0t$, $\beta=0.1t$,
$\Delta_s=0.1t$, $\Delta_t=0.9t$, $L_x=50a_0$, and $L_y=10a_0$.
This parameter choice leads $|N_{\rm ZES}| = N_{\rm clean} = 6$.
The results for a superconductor with clean surface show
twelve ZES as shown with the open symbols.
Although we introduce random impurity potential at its surfaces, 
the flat ZESs remain unchanged as shown with the filled symbols
in agreement with the relation $|N_{\rm ZES}| = N_{\rm clean}$.
This suggests the validity of our conclusion in Sec.~II.
The degeneracy at zero energy is protected by chiral symmetry 
originated from the time-reversal-like symmetry.

%----------------------------------------------------
%-----------------------------------------------------

%%%%%%%%%%%%%%%
\section{Conclusion}
\label{sec:4}
%%%%%%%%%%%%%%%
We have discussed the effects of the random impurity potential on the degenerate zero-energy states
appearing at the surface of a nodal superconductor preserving chiral symmetry.
A method called dimensional reduction enables us to topologically characterize 
nodal superconductors in the presence of translational symmetry.
The number of zero-energy bound states at a clean surface, $N_{\textrm{clean}}$, 
is calculated by using a winding number defined in a one-dimensional Brillouin zone
and is usually much larger than unity proportional to the surface width.
By focusing on the chiral symmetry of the Bogoliubov-de Gennes Hamiltonian, we 
show that an index, $N_{\rm ZES}$, characterizes the number of zero-energy 
states at a dirty surface. 
We confirmed our conclusion with numerical simulations on the tight-binding model.
The index $N_{\rm ZES}$ is defined by the chiral eigenvalues 
of zero-energy states. Simultaneously, $N_{\rm ZES}$ is calculated 
from the winding number in a one-dimensional Brillouin zone. 
The index theorem explains the coincidence of two $N_{\rm ZES}$ calculated in the two different ways.
We conclude that $N_{\rm ZES}$ measures degree of the degeneracy of 
zero-energy states at a dirty superconductor surface.
In experiments, potential disorder is inevitable  in the vicinity of the surface.
Therefore, our conclusion sends a useful message to experimentalists in this field 
as regards choosing a target material.

\begin{acknowledgments}
We are grateful to S. Kobayashi and S. -I. Suzuki for useful discussions.
This work was supported by ``Topological Materials Science'' (No. JP15H05852)
and KAKENHI (Nos. JP26287069 and JP15H03525)
from the Ministry of Education, Culture, Sports, Science and Technology (MEXT)
of Japan and by the Ministry of Education and Science of
the Russian Federation (Grant No. 14Y.26.31.0007).
 SI is supported in part by a Grant-in-Aid for JSPS
Fellows (Grant No. JP16J00956) provided by the Japan Society for the Promotion of Science (JSPS).
\end{acknowledgments}

%%%%%%%%%%%%%%%
\appendix
%%%%%%%%%%%%%%%

%%%%%%%%%%%%%%%
\begin{widetext}
\section{Bogoliubov-de Gennes Hamiltonian on the tight-binding model}
%%%%%%%%%%%%%%%
We present the BdG Hamiltonian on a two-dimensional tight-binding lattice.
The eigenvalues in Sec.~\ref{sec:2} are by diagonalizing the tight-binding 
Hamiltonian. The kinetic energy part and the impurity potential are common for 
all the superconductors and are given by
\begin{align}
H_{\rm kin} =& \sum_{m=1}^{l_y} \sum_{\alpha=\uparrow, \downarrow} \left[
- t \sum_{j=1}^{l_x-1} \left(
 \psi_{\boldsymbol{r}+\hat{\boldsymbol{x}},\alpha}^{\dagger}\psi_{\boldsymbol{r},\alpha} 
 + \psi_{\boldsymbol{r},\alpha}^{\dagger}\psi_{\boldsymbol{r}+\hat{\boldsymbol{x}},\alpha} \right)  
 - t \sum_{j=1}^{l_x}  \left(
\psi_{\boldsymbol{r}+\hat{\boldsymbol{y}},\alpha}^{\dagger}\psi_{\boldsymbol{r},\alpha} 
+ \psi_{\boldsymbol{r},\alpha}^{\dagger}\psi_{\boldsymbol{r}+\hat{\boldsymbol{y}},\alpha} \right)
 + \sum_{j=1}^{l_x} \left(4t - \mu \right) \psi_{\boldsymbol{r},\alpha}^{\dagger}\psi_{\boldsymbol{r},\alpha}\right], \\
H_{\rm imp} =& \left( \sum_{j=1}^{l_d}  + \sum_{j=l_x-l_d+1}^{l_x} \right)\sum_{m=1}^{l_y}
\sum_{\alpha=\uparrow, \downarrow}
v \left( \boldsymbol{r} \right)  \psi_{\boldsymbol{r},\alpha}^{\dagger}\psi_{\boldsymbol{r},\alpha},  
\end{align}
where $\psi_{\boldsymbol{r},\alpha}$ ($\psi_{\boldsymbol{r},\alpha}^\dagger$) is the annihilation (creation) operator of a electron at 
$\boldsymbol{r} = j a_0 \boldsymbol{x} + m a_0 \boldsymbol{y}$ with spin $\alpha$,
$t$ is the hopping integral, and $\mu$ is the chemical potential. 
The unit lattice vector 
in the $x$ and $y$ directions are defined by $\hat{\boldsymbol{x}}$ and $\hat{\boldsymbol{y}}$, respectively.
The number of the lattice site in the $x$ ($y$) direction is represented by $l_x = L_x/a_0$ ($l_y = L_y/a_0$).
We introduce the impurity potential $v \left( \boldsymbol{r} \right)$ in the outermost $l_d = L_d/a_0$ layers in the $x$-direction.
The impurity potential is described by  $v \left( \boldsymbol{r} \right)$
which is given randomly
in the range of $- V_{\rm I}/2 \leq v \left( \boldsymbol{r} \right) \leq V_{\rm I}/2$.

\subsection{Unconventional superconductors}
\label{sec: tb_1}
The total Hamiltonian of the unconventional superconductors
is represented as $H = H_{\rm kin} + H_{\mu}  + H_{\rm imp}$, 
where $H_\mu$ for $\mu=p$, $d$ and $f$ depends on the pairing symmetry as
\begin{align}
H_p =&  \sum_{j=1}^{l_x-1} \sum_{m=1}^{l_y} \sum_{\alpha} 
\frac{ i \Delta_0}{2}  \left(  \psi_{\boldsymbol{r}+\hat{\boldsymbol{x}},\alpha}^{\dagger}
\psi_{\boldsymbol{r},\bar{\alpha}}^{\dagger}
-  \psi_{\boldsymbol{r},\alpha}^{\dagger}
\psi_{\boldsymbol{r}+\hat{\boldsymbol{x}},\bar{\alpha}}^{\dagger} \right) \;+  {\rm H.c.}, \\
H_d =&  \sum_{j=1}^{l_x-1} \sum_{m=1}^{l_y} \sum_{\alpha} 
\frac{ s_\alpha \Delta_0}{4}  \left(  
\psi_{\boldsymbol{r}+\hat{\boldsymbol{x}}+\hat{\boldsymbol{y}},\alpha}^{\dagger}
 \psi_{\boldsymbol{r},\bar{\alpha}}^{\dagger} 
+  \psi_{\boldsymbol{r},\alpha}^{\dagger}
 \psi_{\boldsymbol{r}+\hat{\boldsymbol{x}}+\hat{\boldsymbol{y}},\bar{\alpha}}^{\dagger}
- \psi_{\boldsymbol{r}+\hat{\boldsymbol{x}}-\hat{\boldsymbol{y}},\alpha}^{\dagger}
 \psi_{\boldsymbol{r},\bar{\alpha}}^{\dagger} 
-  \psi_{\boldsymbol{r},\alpha}^{\dagger}
\psi_{\boldsymbol{r}+\hat{\boldsymbol{x}}-\hat{\boldsymbol{y}},\bar{\alpha}}^{\dagger}
 \right)+  {\rm H.c.},\\
H_f =&  \sum_{j=1}^{l_x-1} \sum_{m=1}^{l_y} \sum_{\alpha}  \frac{i \Delta_0}{4} 
 \left( \psi_{\boldsymbol{r}+\hat{\boldsymbol{x}}+2\hat{\boldsymbol{y}},\alpha}^{\dagger} 
 \psi_{\boldsymbol{r},\bar{\alpha}}^{\dagger}
+  \psi_{\boldsymbol{r}+\hat{\boldsymbol{x}}-2\hat{\boldsymbol{y}},\alpha}^{\dagger} 
\psi_{\boldsymbol{r},\bar{\alpha}}^{\dagger} 
-  \psi_{\boldsymbol{r},\alpha }^{\dagger} 
\psi_{\boldsymbol{r}+\hat{\boldsymbol{x}}+2\hat{\boldsymbol{y}},\bar{\alpha}}^{\dagger} 
-  \psi_{\boldsymbol{r},\alpha}^{\dagger} 
\psi_{\boldsymbol{r}+\hat{\boldsymbol{x}}-2\hat{\boldsymbol{y}},\bar{\alpha}}^{\dagger} \right) +  {\rm H.c.},
\end{align}
where $\bar{\alpha}$ is the opposite spin of $\alpha$ and $\Delta_0$ is the pair potential at zero temperature.
The factor$s_\alpha$ is  $+1$ for $\alpha=\uparrow$ and is $-1$ for $\alpha=\downarrow$.
For spin-triplet case, we assume a Cooper pair consists of two electrons with the opposite spin directions.
In Sec.~\ref{sec:2}, we diagonalize the reduced BdG Hamiltonian into $2\times 2$ Nambu space.

\subsection{$\boldsymbol{(d_{xy} + p)}$-wave superconductor}
\label{sec: tb_2}
The Hamiltonian of the $(d_{xy} + p)$-wave superconductor
discussed in Sec.~\ref{sec:2}B is described by 
adding the Rashba spin orbit interaction $H_{\rm R}$ and the pair potential $H_{d_{xy} + p}$ to 
$H_{\rm kin} + H_{\rm imp}$.
The spin-orbit coupling term is represented by 
\begin{align}
H_{\rm R} =& - i \frac{\lambda_{\rm R}}{2} \sum_{\alpha,\beta}   \sum_{m=1}^{l_y} 
\left[ \sum_{j=1}^{l_x-1} (\sigma_{y})_{\alpha,\beta} \left(
\psi_{\boldsymbol{r}+\hat{\boldsymbol{x}},\alpha}^{\dagger} \psi_{\boldsymbol{r}, \beta} -  
\psi_{\boldsymbol{r}, \alpha}^{\dagger} \psi_{\boldsymbol{r}+\hat{\boldsymbol{x}}, \beta} \right)
 - \sum_{j=1}^{l_x} (\sigma_{x})_{\alpha,\beta} \left(
\psi_{\boldsymbol{r}+\hat{\boldsymbol{y}},\alpha}^{\dagger} \psi_{\boldsymbol{r}, \beta}  - 
\psi_{\boldsymbol{r},\alpha}^{\dagger}\psi_{\boldsymbol{r}+\hat{\boldsymbol{y}},\beta} \right) 
\right].
\end{align}
The pair potential consists of five parts:
$H_{d_{xy} + p} = H_{\Delta 1} + H_{\Delta 1} +H_{\Delta 2} +H_{\Delta 4} +H_{\Delta 5}$.
The each parts are
represented by
\begin{align}
H_{\Delta 1} =& i \frac{\Delta_s}{4} \sum_{\alpha,\beta}  
\sum_{j=1}^{l_x-1} \sum_{m=1}^{l_y} (\sigma_{y})_{\alpha,\beta}  
\left[  
\psi_{\boldsymbol{r}+\hat{\boldsymbol{x}}+\hat{\boldsymbol{y}},\alpha}^{\dagger} 
\psi_{\boldsymbol{r},\beta}^{\dagger} +  
\psi_{\boldsymbol{r},\alpha}^{\dagger} 
\psi_{\boldsymbol{r}+\hat{\boldsymbol{x}}+\hat{\boldsymbol{y}}, \beta}^{\dagger}
 -\psi_{\boldsymbol{r}+\hat{\boldsymbol{x}}-\hat{\boldsymbol{y}},\alpha}^{\dagger}
 \psi_{\boldsymbol{r},\beta}^{\dagger}  - 
\psi_{\boldsymbol{r},\alpha}^{\dagger}\psi_{\boldsymbol{r}+\hat{\boldsymbol{x}}-\hat{\boldsymbol{y}}, \beta}^{\dagger} 
\right]+ {\rm H.c.},\\
H_{\Delta 2} =& - i \frac{\Delta_t}{4} \sum_{\alpha,\beta}  \sum_{j=1}^{l_x-1} 
\sum_{m=1}^{l_y} (\sigma_{z})_{\alpha,\beta}  \left[  
\psi_{\boldsymbol{r}+\hat{\boldsymbol{x}},\alpha}^{\dagger} 
\psi_{\boldsymbol{r},\beta}^{\dagger} 
-  
\psi_{\boldsymbol{r},\alpha}^{\dagger} 
\psi_{\boldsymbol{r}+\hat{\boldsymbol{x}},\beta}^{\dagger}
\right]+ {\rm H.c.}, \\
H_{\Delta 3} =& \frac{\Delta_t}{4} \sum_{\alpha}  \sum_{j=1}^{l_x} 
\sum_{m=1}^{l_y}   \left[  
\psi_{\boldsymbol{r}+\hat{\boldsymbol{y}},\alpha}^{\dagger} 
\psi_{\boldsymbol{r},\alpha}^{\dagger} 
-  
\psi_{\boldsymbol{r},\alpha}^{\dagger} 
\psi_{\boldsymbol{r}+\hat{\boldsymbol{y}},\alpha}^{\dagger}
\right]+ {\rm H.c.}, \\
H_{\Delta 4} =&  i \frac{\Delta_t}{8} \sum_{\alpha,\beta}  \sum_{j=1}^{l_x-1} \sum_{m=1}^{l_y} (\sigma_{z})_{\alpha,\beta}
\left[  
\psi_{\boldsymbol{r}+\hat{\boldsymbol{x}}+2\hat{\boldsymbol{y}},\alpha}^{\dagger} 
 \psi_{\boldsymbol{r},\beta}^{\dagger}
+  \psi_{\boldsymbol{r}+\hat{\boldsymbol{x}}-2\hat{\boldsymbol{y}},\alpha}^{\dagger} 
\psi_{\boldsymbol{r},\beta}^{\dagger} 
-  \psi_{\boldsymbol{r},\alpha }^{\dagger} 
\psi_{\boldsymbol{r}+\hat{\boldsymbol{x}}+2\hat{\boldsymbol{y}},\beta}^{\dagger} 
-  \psi_{\boldsymbol{r},\alpha}^{\dagger} 
\psi_{\boldsymbol{r}+\hat{\boldsymbol{x}}-2\hat{\boldsymbol{y}},\beta}^{\dagger}
\right]+ {\rm H.c.},\\
H_{\Delta 5} =&  - \frac{\Delta_t}{8} \sum_{\alpha}  \sum_{j=1}^{l_x-2} \sum_{m=1}^{l_y}
\left[  
\psi_{\boldsymbol{r}+2\hat{\boldsymbol{x}}+\hat{\boldsymbol{y}},\alpha}^{\dagger} 
 \psi_{\boldsymbol{r},\alpha}^{\dagger}
+  \psi_{\boldsymbol{r}-2\hat{\boldsymbol{x}}+\hat{\boldsymbol{y}},\alpha}^{\dagger} 
\psi_{\boldsymbol{r},\alpha}^{\dagger} 
-  \psi_{\boldsymbol{r},\alpha }^{\dagger} 
\psi_{\boldsymbol{r}+2\hat{\boldsymbol{x}}+\hat{\boldsymbol{y}},\alpha}^{\dagger} 
-  \psi_{\boldsymbol{r},\alpha}^{\dagger} 
\psi_{\boldsymbol{r}-2\hat{\boldsymbol{x}}+\hat{\boldsymbol{y}},\alpha}^{\dagger}
\right]+ {\rm H.c.}
\end{align}
The amplitude of the pair potential for the spin-singlet (-triplet) component is represented by $\Delta_s$ ($\Delta_t$).
The Pauli matrices in spin space are represented by $\sigma_{\nu}$ ($\nu=x,y,z$).

\subsection{Noncentrosymmetric superconductor with the persistent helix states}
\label{sec: tb_3}
The BdG Hamiltonian for a NCS with the persistent spin-helix states discussed in Sec.~\ref{sec:3b}
is described by $H =H_{\rm kin} + H_{\rm D}^{110} + H_{\Delta_p}  + H_{\rm imp}$.
The spin-orbit coupling and the pair potential are given by
\begin{align}
H_{\rm D}^{110} =&  i \frac{\lambda_{D}}{2} \sum_{\alpha,\beta}  \sum_{j=1}^{l_x-1} \sum_{m=1}^{l_y} 
(\sigma_{z})_{\alpha,\beta} \left(
\psi_{\boldsymbol{r}+\hat{\boldsymbol{x}},\alpha}^{\dagger} \psi_{\boldsymbol{r},\beta}  -  
\psi_{\boldsymbol{r},\alpha}^{\dagger} \psi_{\boldsymbol{r}+\hat{\boldsymbol{x}},\beta} \right),\\
H_{\Delta_p} =&  \sum_{\alpha,\beta} \sum_{m=1}^{l_y}
\left[
i \Delta_s 
\sum_{j=1}^{l_x}
(\sigma_{y})_{\alpha,\beta} 
\psi_{\boldsymbol{r},\alpha}^{\dagger} \psi_{\boldsymbol{r},\beta}^{\dagger} 
 + i \frac{\Delta_t}{2} \sum_{j=1}^{l_x-1} 
(\sigma_{x})_{\alpha,\beta} 
\left(
\psi_{\boldsymbol{r}+\hat{\boldsymbol{x}},\alpha}^{\dagger} \psi_{\boldsymbol{r},\beta}^{\dagger}
  -  
\psi_{\boldsymbol{r},\alpha}^{\dagger} \psi_{\boldsymbol{r}+\hat{\boldsymbol{x}},\beta}^{\dagger}  \right) 
\right]+ {\rm H.c.},
\end{align}
\end{widetext}
%where $\lambda_{\rm D}$ denotes the strength of the Dresselhaus[110] spin-orbit coupling.

%%%%%%%%%%%%%%%
\section{Unitary transformation for the $\boldsymbol{(d_{xy} + p)}$-wave superconductor}
%%%%%%%%%%%%%%%
\label{sec:uni1}
%%%%%%%%%%%%%%%
The BdG Hamiltonian of a $(d_{xy} + p)$ superconductor in Sec.~\ref{sec:2} B is represented by
\begin{align}
&H_{\boldsymbol{k}} = \left[
\begin{array}{cc}
\hat{h}_{\boldsymbol{k}} & \hat{\Delta}_{\boldsymbol{k}} \\
- \hat{\Delta}_{-\boldsymbol{k}}^{\ast} & -\hat{h}_{-\boldsymbol{k}}^{\ast} \\
\end{array}\right],\\
&\hat{h}_{\boldsymbol{k}}=\xi(\boldsymbol{k}) \hat{\sigma}_0
+ \boldsymbol{g}_{\rm r}(\boldsymbol{k}) \cdot \hat{\boldsymbol{\sigma}},\\
&\hat{\Delta}_{\boldsymbol{k}}=i f (\boldsymbol{k})[
\Delta_s +\Delta_t \frac{\Delta_t}{\alpha k} \boldsymbol{g}_{\rm r}(\boldsymbol{k})\cdot \hat{\boldsymbol{\sigma}} ] \hat{\sigma}_2,
\end{align}
where $\boldsymbol{g}_{\rm r}(\boldsymbol{k}) = \alpha( k_y \boldsymbol{x} - k_x \boldsymbol{y})$,
 $f(\boldsymbol{k}) = (k_x k_y/k^2)$ and $k = \sqrt{k_x^2 + k_y^2 }$.
We first apply a unitary transformation to $U_{\boldsymbol{k}}^{\dagger} H_{\boldsymbol{k}} U_{\boldsymbol{k}} = H_{\boldsymbol{k}}^{\prime}$ with
\begin{align}
%&U_{\boldsymbol{k}}^{\dagger} H_{\boldsymbol{k}} U_{\boldsymbol{k}} = H_{\boldsymbol{k}}^{\prime},\\
%
&U_{\boldsymbol{k}} = \left[
\begin{array}{cc}
\hat{u}_{\boldsymbol{k}} & 0 \\
0 &  \hat{u}_{-\boldsymbol{k}}^{\ast} \\
\end{array}\right], \\
&\hat{u}_{\boldsymbol{k}}= \frac{1}{\sqrt{2}}\left[
\begin{array}{cc}
1 & ie^{- i \theta_{\boldsymbol{k}}} \\
-i e^{i \theta_{\boldsymbol{k}}} &  -1\\
\end{array}\right],\quad
\theta_{\boldsymbol{k}} = \arctan \left[ \frac{k_y }{k_x} \right].
\end{align}
The second unitary transformation 
\begin{align}
H_{\gamma}(\boldsymbol{k}) = U_{0}^\dagger H_{\boldsymbol{k}}^{\prime} U_{0}, \quad
U_{0}=\left[ \begin{array}{cccc}
1 & 0 & 0 & 0\\
0 & 0 & 1 & 0 \\
0 & 0 & 0 & 1 \\
0 & 1 & 0 & 0
\end{array}\right],
\end{align}
results in 
\begin{align}
&H_{\gamma}(\boldsymbol{k})= \left[
\begin{array}{cc}
\hat{H}_+ (\boldsymbol{k}) & 0 \\
0 & \hat{H}_- (\boldsymbol{k}) \\
\end{array}\right],\\
&\hat{H}_{\pm}(\boldsymbol{k})= \left[
\begin{array}{cc}
E_{\pm} (\boldsymbol{k}) & -\Delta_{\pm}(\boldsymbol{k})\\
- \Delta_{\pm}(\boldsymbol{k}) &  - E_{\pm}(\boldsymbol{k}) \\
\end{array}\right],\\
&E_{\pm}(\boldsymbol{k}) = \xi(\boldsymbol{k}) \pm  \mid \boldsymbol{g}_{\rm R}(\boldsymbol{k})\mid,\\
&\Delta_{\pm} (\boldsymbol{k}) = f(\boldsymbol{k}) [ \Delta_t \pm \Delta_s].
\end{align}

%%%%%%%%%%%%%%%%%%
\section{Persistent spin-helix states with coexistence
of Rashba and Dresselhaus[100] spin-orbit coupling}
\label{sec:psh_rd}
%%%%%%%%%%%%%%%%%%
We  shown an alternative way to realize the persistent spin-helix states~\cite{ bernevig,chang}.
Let us consider the thin film growing along the [001] crystal direction. 
In such the two dimensional eletron system, the Rashba type SOC
$\boldsymbol{g}_{\rm r}(\boldsymbol{k}) = \alpha( k_y \boldsymbol{x} - k_x \boldsymbol{y})$
and the Dresselhaus [001] type SOC
$\boldsymbol{g}_{\rm d}^{\prime}(\boldsymbol{k}) = \beta^{\prime}( k_x \boldsymbol{x} - k_y \boldsymbol{y})$ coexist.
The Hamiltonian is described as
\begin{align}
&\hat{h}_{\rm RD} (\boldsymbol{k})
= \xi(\boldsymbol{k}) \hat{\sigma}_0 + \hat{h}_{\rm R} (\boldsymbol{k}) + \hat{h}_{\rm D}^{100} (\boldsymbol{k}),\\
&\hat{h}_{R} = \alpha\left( k_{y} \hat{\sigma}_{x} - k_{x} \hat{\sigma}_{y} \right), \\
&\hat{h}_{D}^{100} = \beta^{\prime} \left( k_{x} \hat{\sigma}_{x} - k_{y} \hat{\sigma}_{y} \right).
\end{align}
When we define
\begin{align}
&k_{\pm} = \frac{1}{\sqrt{2}} \left( k_{x} \pm k_{y} \right), 
\end{align}
the Hamiltonian is rewritten as
\begin{align}
\hat{h}_{\rm RD} (\boldsymbol{k})=\xi^{\prime}(\boldsymbol{k}) \hat{\sigma}_0
+ \hat{h}_{+} + \hat{h}_{-}, 
\end{align}
where
\begin{align}
&\xi^{\prime}(\boldsymbol{k}) =\frac{\hbar^2}{2m} \left( k_{+}^{2} + k_{-}^{2} \right)
 - \mu_{\rm F},\\
&\hat{h}_{\pm} =  \lambda_{\pm} k_{\pm} \hat{\sigma}_{\pm}, \\
&\lambda_{\pm} = \frac{1}{\hbar} \left( \beta \pm \alpha \right),
\hspace{10pt}
\hat{\sigma}_{\pm} = \frac{1}{\sqrt{2}}\left( \hat{\sigma}_{x} \mp \hat{\sigma}_{y} \right).
\end{align}
The strength of the Rashba SOC is tunable by an externally applied electric field.
When we consider a special case of $\alpha=\beta^{\prime}$,
which can be experimentally accessible~\cite{kohda,salis},
the Hamiltonian is deformed as
\begin{align}
\hat{h} (\boldsymbol{k})
= \left[ \frac{\hbar^2 } {2m} (k_+^2 + k_-^2) - \mu_{\rm F}
\right] \hat{\sigma}_0
+ \beta^{\prime} p_+ \hat{\sigma}_+.
\label{eq:psh2}
\end{align}
The Hamiltonian in Eq.~(\ref{eq:psh2}) is unitary equivalent to that in Eq.~(\ref{eq:psh})~\cite{ bernevig}.
Therefore the persistent spin-helix states can also obtained in  the thin film growing along the [001] crystal direction.
Moreover, as discussed in Sec.~\ref{sec:3b},
we can expect the flat ZESs at a dirty surface of a
superconducting thin film with coexistence of Rashba and Dresselhaus[001] spin-orbit coupling.

%%%%%%%%%%%%%%%

\end{document}